\documentclass[fleqn,usenatbib]{mnras}

\usepackage{newtxtext,newtxmath}

\usepackage[T1]{fontenc}

\DeclareRobustCommand{\VAN}[3]{#2}
\let\VANthebibliography\thebibliography
\def\thebibliography{\DeclareRobustCommand{\VAN}[3]{##3}\VANthebibliography}

\usepackage{graphicx}	
\usepackage{amsmath}	

\title[Coverage Dependent H$_2$ Desorption Energy]{Coverage Dependent H$_2$ Desorption Energy : a Quantitative Explanation Based on Encounter Desorption Mechanism}

\author[Q. Meng et al.]{Qingkuan Meng,$^{1}$
Qiang Chang,$^{1}$\thanks{E-mail: changqiang@sdut.edu.cn}
Gang Zhao,$^{1}$
Donghui Quan,$^{2,3}$
Masashi Tsuge,$^{4}$
Xia Zhang,$^{3}$
\newauthor
Yong Zhang$^{5}$
and Xiao-Hu Li$^{3}$
\\
$^{1}$School of Physics and Optoeletronic Engineering, Shandong University of Technology, Zibo 255000, China\\
$^{2}$Research Center for Intelligent Computing Platforms, Zhejiang Laboratory, Hangzhou 311100, China\\
$^{3}$Xinjiang Astronomical Observatory, Chinese Academy of Sciences, 150 Science 1-Street, Urumqi 830011, China\\
$^{4}$Institute of Low Temperature Science, Hokkaido University, Kita-19, Nishi-8, Kita-ku, Sapporo 060-0819, Japan\\
$^{5}$School of Physics and Astronomy, Sun Yat-sen University, Zhuhai 519082, China}

\date{Accepted XXX. Received YYY; in original form ZZZ}

\pubyear{2015}

\begin{document}
\label{firstpage}
\pagerange{\pageref{firstpage}--\pageref{lastpage}}
\maketitle

\begin{abstract}
Recent experiments show that the desorption energy of H$_2$ on a diamond-like carbon (DLC) surface depends on the H$_2$ coverage of the surface. We aim to quantitatively explain the coverage dependent H$_2$ desorption energy measured by the experiments. We derive a math formula to calculate an effective H$_2$ desorption energy based on the encounter desorption mechanism. The effective H$_2$ desorption energy depends on two key parameters, the desorption energy of H$_2$ on H$_2$ substrate and the ratio of H$_2$ diffusion barrier to its desorption energy. The calculated effective H$_2$ desorption energy qualitatively agrees with the coverage dependent H$_2$ desorption energy measured by the experiments if the values of these two parameters in literature are used in the calculations. We argue that the difference between the effective H$_2$ desorption energy and the experimental results is due to the lacking of knowledge about these
two parameters. So, we recalculate these two parameters based on experimental data. Good agreement between theoretical and experimental results can be achieved if these two updated parameters are used in the calculations. 
\end{abstract}




\section{Introduction}
Not all interstellar molecules can be efficiently formed in the gas-phase. Species such as molecular hydrogen or methanol were believed to be mainly formed on dust grains~\citep{Gould1963,Watanabe2002,Wakelam2017,Tsuge2023}. The desorption energies of surface species play essential roles in the synthesis of species on interstellar dust grains. This parameter determines the residence time of surface species on grains. Because the time period that species reside on grains must be long enough to find a partner species to react, their desorption energies must be large enough in order that these species participate in surface chemical reactions. On the other hand, because the diffusion barriers of species are believed to be proportional to their desorption energies~\citep{Hasegawa1992,Herbst2009,Garrod2008}, their desorption energies must be small enough so that they can efficiently diffuse to react with other species on grain surfaces.

The desorption energies of species are usually assumed to be independent of their population on grains in astrochemical models. However, theoretical and experimental studies in the past few decades showed that the desorption energies of species are usually coverage dependent~\citep{Yates1979,Wong2019}. Moreover, the desorption energies typically decrease with the coverage, which implies repulsive interactions between nearest neighbors~\citep{Wong2019}. While most of these studies are not relevant to the field of astrochemistry, \citet{Tsuge2019} recently studied the coverage dependent desorption energy of H$_2$ on a diamond-like carbon surface, which may be analogous to interstellar dusts. They found that the H$_2$ desorption energy drops as the coverage of surface H$_2$ increases. However, \citet{Tsuge2019} did not explain the reason for their findings.   

To the best of our knowledge, the coverage dependent H$_2$ desorption energy has not been adopted in astrochemical models, although recent studies suggest that the existence of surface H$_2$ should decrease H$_2$ desorption energy on icy grain surfaces~\citep{Hincelin2015,Garrod2011}. If we only use the desorption energy of H$_2$ on water ice in chemical models,  almost all gas-phase H$_2$ molecules would be frozen on grain surfaces if the grain temperature is low ($\sim 10 $ K) and the gas number density is high ( $\ge10^{12}$ cm$^{-3}$)~\citep{Hincelin2015},  which has not been observed so far. Therefore, astrochemical models usually assume a much reduced desorption of H$_2$ on sites already occupied by another H$_2$ to increase the desorption rate of surface H$_2$~\citep{Hincelin2015,Garrod2011}.  

A convenient way to incorporate the much reduced H$_2$ desorption energy on H$_2$ substrates in astrochemical models is to use the encounter desorption (ED) mechanism~\citep{Hincelin2015,Chang2021,Zhao2022}. 
This mechanism assumes that when a H atom or H$_2$ molecule encounters a H$_2$ molecule, its desorption energy immediately drops from the energy value on water ice surface to that on H$_2$ substrates. To include ED mechanism in rate equation models, we only have to add an reaction JH$_2$ + JH$_2$ $\rightarrow$ JH$_2$ + H$_2$ or JH + JH$_2$ $\rightarrow$ H + JH$_2$ in the surface reaction network, where letter J designates surface species. Rate equation chemical models that include the ED mechanism predict that the abundance of JH$_2$ molecules is only a few monolayers~\citep{Hincelin2015}, thus, these model results are not in conflict with observations.

We can use the ED mechanism to intuitively explain the coverage dependent H$_2$ desorption energy. As the coverage of JH$_2$ increases, JH$_2$ molecules can encounter other JH$_2$ molecules more frequently. So, JH$_2$ molecules are more likely to desorb as the population of JH$_2$ increases. As a result, the JH$_2$ desorption energy measured by experiments should decrease as the coverage of JH$_2$ increases.

In this work, our purpose is to quantitatively explain the coverage dependent JH$_2$ desorption energy. We derive a formula to calculate the coverage dependent H$_2$ desorption energy based on the ED mechanism and then compare with experimental results. We think that the discrepancy between our calculated coverage dependent JH$_2$ desorption energies and these measured by \citet{Tsuge2019} is because of the poor knowledge about two parameters, which are the diffuse barrier of JH$_2$ and the desorption energy of H$_2$ molecules when they encounter with each other. We recalculate these two parameters based on experimental data for better agreement between theoretical and experimental results.

\section{Methods}
Based on the ED mechanism, JH$_2$ molecules desorb via two ways in the experiments performed by~\citet{Tsuge2019}. The first one is the thermal desorption on DLC sites while the second one is the encounter desorption of JH$_2$ molecules when two hydrogen molecules meet in the same site. The thermal desorption rate coefficient is,
\begin{equation}
	k_{D1} = \nu \exp(-E_{D_{DLC}}/kT), 
\end{equation}
where $E_{D_{DLC}}$, $\nu$, $k$ and $T$ are the desorption energy of JH$_2$ on DLC sites, the attempt frequency,
the Boltzmann constant and the surface temperature respectively.
In this paper, energy values are usually given in unit of K. However, if an energy value was given in unit of meV in the reference, 
we also express it in meV. Energy values expressed in K can be converted to that in meV by multiplying the Boltzmann constant, k. 
 We assumed $15 \le T \le 25$ K because the temperature-programmed desorption  (TPD) signal is too weak to be distinguished from the background noise at $T$ that is out of the temperature range in the experiments. \citet{Tsuge2019} argued that $E_{D_{DLC}}$ should be approximately the same as the activation energy for H$_2$ desorption from a graphite  surface (41 meV) although what they measured is well  below 41 meV (476 K) at relative coverage above 0.04. 
Moreover, they found that the coverage dependent JH$_2$ desorption energy can rise as high as 40 meV (464 K) at relative 
coverage below 0.04. Because $E_{D_{DLC}}$ is actually the coverage dependent JH$_2$ desorption energy when the coverage is so low that no encounter desorption events occur on the surface, we assumed $E_{D_{DLC}}=41$ meV (476 K) in this work. The JH$_2$ encounter desorption rate is~\citep{Hincelin2015},
\begin{equation}
R_{D2} = \frac{1}{2}N_{H_2}^2 k_{enc} P_{des},
\end{equation}
where $N_{H_2}$ is the population of JH$_2$, $k_{enc}$ is the rate coefficient for a JH$_2$ molecule to encounter another one while $P_{des}$ is the probability that JH$_2$ molecules desorb when the encounter events occur. \citet{Hincelin2015} 
found that $k_{enc}$ can be calculated as
\begin{equation} 
	k_{enc} =\frac{2\nu}{N_s} \exp(-E_{b_{DLC}}/kT),
\end{equation}  
where $N_s$ is the number of sites on surfaces while $E_{b_{DLC}}$ is the diffusion barrier of JH$_2$ on the DLC binding sites. However, the value of $E_{b_{DLC}}$ is not well known. In astrochemical models, the diffusion barrier of a species is usually chosen to be a ratio of its desorption energy. Two ratios, 0.3 or 0.5 were usually used in full gas-grain reaction network astrochemical models~\citep{Hasegawa1992,Herbst2009,Garrod2008}. On the other hand, the ratio can be as high as 0.77 in the molecular hydrogen formation models~\citep{Chang2005,Zhao2022}.
Therefore, we varied and studied $\alpha = E_{b_{DLC}}/E_{D_{DLC}} $ = 0.77, 0.5 and 0.3 in this work.
The probability $P_{des}$ can be calculated as, 
\begin{equation} 
		P_{des} = \frac{\exp(-E_{D H_{2}}/kT)} {\exp(-E_{D H_{2}}/kT)+\exp(-E_{b H_{2}}/kT)},
\label{equ1}
\end{equation} 
where $E_{D H_{2}}$ and $E_{b H_{2}}$ are the desorption energy and diffusion barrier of JH$_2$ on sites occupied by JH$_2$ respectively. The value of $E_{D H_{2}}$ is not well known either. To the best of our knowledge, $E_{D H_{2}}$ has not been measured in laboratories. So, we used two theoretical values in this work. The first one is 23 K, which was suggested by \citet{Cuppen2007}. Based on more rigorous quantum chemical calculations, \citet{Das2021} suggested $E_{D H_{2}}$ is between 67 and 79 K. So, the second value is set to be the intermediate value, 73 K as in \citet{Zhao2022}. We used a scaling equation, $E_{b H_{2}}$ = $\alpha E_{D H_{2}}$ to calculate $E_{b H_{2}}$. The total JH$_2$ desorption rate, $R_{T}$, is, 
\begin{equation}
R_{T} = k_{D1} N_{H_2} + \frac{1}{2}N_{H_2}^2 k_{enc} P_{des}.
\end{equation}

We assumed an effective desorption coefficient $k_{eva}$~\citep{Chang2006}, so that $R_{T}=k_{eva} N_{H_2}$. We can interpret $k_{eva}$ as the average JH$_2$ desorption rate coefficient on all sites. We derived an effective desorption energy, 
$E_{effD}$ by the following equation,
\begin{equation}
	\begin{array}{lll} 
			\nu \exp(-E_{effD}/kT) &  =  & k_{eva} \\
	                      &  =  & k_{D1} +  \frac{1}{2}N_{H_2} k_{enc} P_{des}. \\
\end{array}
\end{equation} 
So, $E_{effD}$ can be calculated as,
\begin{equation}
\resizebox{0.9\hsize}{!}{$
\begin{array}{llll} 
		E_{effD} & = & -kT ln((k_{D1} +  \frac{1}{2}N_{H_2} k_{enc} P_{des})/\nu) \\
		& = &  -kT ln(exp(-E_{D_{DLC}} /kT) + exp(-E_{b_{DLC}}/kT) \theta P_{des}),\\
	& = & E_{D_{DLC}} -kT ln(1+exp(( E_{D_{DLC}} - E_{b_{DLC}} )/kT) \theta P_{des}), \\
\end{array}$}
\label{equ2}
\end{equation} 
where $\theta$ is the relative coverage, $N_{H_2}/N_s$. 
The coverage dependent nature of $E_{effD}$ is obvious in equation~(\ref{equ2}). Moreover, we can see that $E_{effD}$ is a function of $T$, but no longer depends on $\nu$. We will compare the derived $E_{effD}$ with the coverage dependent H$_2$ desorption energy measured by~\citet{Tsuge2019} in the next section. 

In addition to reporting the coverage dependent H$_2$ desorption energy, $E_{exptD}$, determined using the complete analysis method~\citep{King1975}, \citet{Tsuge2019} calculated another set of coverage dependent H$_2$ desorption energy, $E_{exptD2}$ based on their TPD spectrum and the Polanyi-Wigner equation. They found that $E_{exptD2}$ agrees well with $E_{exptD}$. Because the variation in $E_{exptD}$ is much larger than that in $E_{exptD2}$, we compare $E_{effD}$ with $E_{exptD2}$ in this work. If the difference between $E_{effD}$ and $E_{exptD2}$ is smaller than the variation in $E_{exptD}$, we conclude that $E_{effD}$ agrees reasonably well with both $E_{exptD2}$ and $E_{exptD}$. 

In order for $E_{effD}$ to agree better with the coverage dependent H$_2$ desorption energy measured in laboratories, the values of $E_{D H_{2}}$ or $\alpha$ may be different than these in the literature. As mentioned earlier, the values of these two parameters in the literature are no better than theoretical predictions. 

If $E_{D {H_{2}}_b}$ and $\alpha_b$ are the "true" values of $E_{D H_{2}}$ and $\alpha$, respectively, that reproduce the experimental results, they should satisfy the following equation,
\begin{equation}
E_{effD}(\theta, E_{D {H_{2}}_b}, \alpha_b, T) = E_{exptD2}(\theta).
\label{equ3}
\end{equation}
Equation~(\ref{equ3}) has two variables, $E_{D {H_{2}}_b}$ and $\alpha_b$. Because it is not possible to solve a single equation with two variables, we assumed $\alpha_b=0.77, 0.5$ or 0.3 to calculate $E_{D {H_{2}}_b}$. We calculated $E_{D {H_{2}}_b}$ at five surface temperatures, 15 K, 17.5 K, 20 K, 22.5 K and 25 K. Alternatively, we assumed $E_{D {H_{2}}_b} = 23$ or 73 K to calculate $\alpha_b$ at these five temperatures. Essentially, we assumed the value of one parameter ($E_{D H_{2}}$ or $\alpha$) in the literature is true and then calculated the other one. 

We numerically solved equation~(\ref{equ3}) and calculated $E_{D {H_{2}}_b}$ for each $E_{exptD2}(\theta)$ data point with $\alpha_b=0.77, 0.5$ and 0.3 in the calculations. The obtained $E_{D {H_{2}}_b}$ varies according to  $\theta$, so we calculated the average of $E_{D {H_{2}}_b}$, 
$\bar{E}_{D {H_{2}}_b}$ using the following equation,
\begin{equation}
\int_{a}^{b} E_{D {H_{2}}_b} (\theta) d \theta = (b-a) \bar{E}_{D {H_{2}}_b},
\label{equ4}
\end{equation}
where $a$ and $b$ are the minimum and maximum relative coverage respectively. 
We set $a~=~0.03$ and $b~=~0.38$ because the minimum and maximum $\theta$ in the experimental 
data points $E_{exptD}(\theta)$ are 0.03 and 0.38 respectively. Because $E_{D {H_{2}}_b}$ varies according to $T$, the values of $\bar{E}_{D {H_{2}}_b}$ are also different at different temperatures. 

To calculate $E_{effD}$ , we set $E_{D {H_{2}}}= \bar{E}_{D {H_{2}}_b}$. We calculated $E_{effD}$ using equation~(\ref{equ2}) and then compared with experimental results. We set $T$ = 15 K, 20 K and 25 K in the calculations of $E_{effD}$. The values of $E_{D {H_{2}}}$ were set to be the values of $\bar{E}_{D {H_{2}}_b}$ at 15 K, 20 K and 25 K respectively in the calculations.

We also studied how well $E_{effD}$ agree with experimental results if a temperature independent value of $E_{D {H_{2}}}$ is used to calculate $E_{effD}$. Because the peak TPD signal occurs at around $T$ = 20 K, we fixed $E_{D {H_{2}}}$ to be the value of $\bar{E}_{D {H_{2}}_b}$ at 20 K and calculated $E_{effD}$ at $T$ = 15 K and 25 K respectively. The obtained $E_{effD}$ were also compared with experimental results.

Similarly, we set $E_{D {H_{2}}_b} = 23$ K and 73 K respectively and solved equation~(\ref{equ3}) to calculate $\alpha_b$. The average of $\alpha_b(\theta)$, $\bar{\alpha}_b$ was calculated using the following equation,
\begin{equation}
\int_{a}^{b}  \alpha_b(\theta) d \theta = (b-a) \bar{\alpha}_b.
\label{equ5}
\end{equation}
We set $\alpha=\bar{\alpha}_b$ and calculated $E_{effD}$ at $T$ = 15 K and 25 K respectively using equation~(\ref{equ2}). We performed two types of calculations. We set $\alpha$ to be the $\bar{\alpha}_b$ values at 15 K and 25 K respectively for the first one while $\alpha$ was fixed to be the value of $\bar{\alpha}_b$ at 20 K in the second one. The obtained $E_{effD}$ were also compared with experimental results.

\section{Results}
\subsection{The effective H$_2$ desorption energy}

\begin{figure}
\centering
\resizebox{8.5cm}{6.8cm}{\includegraphics{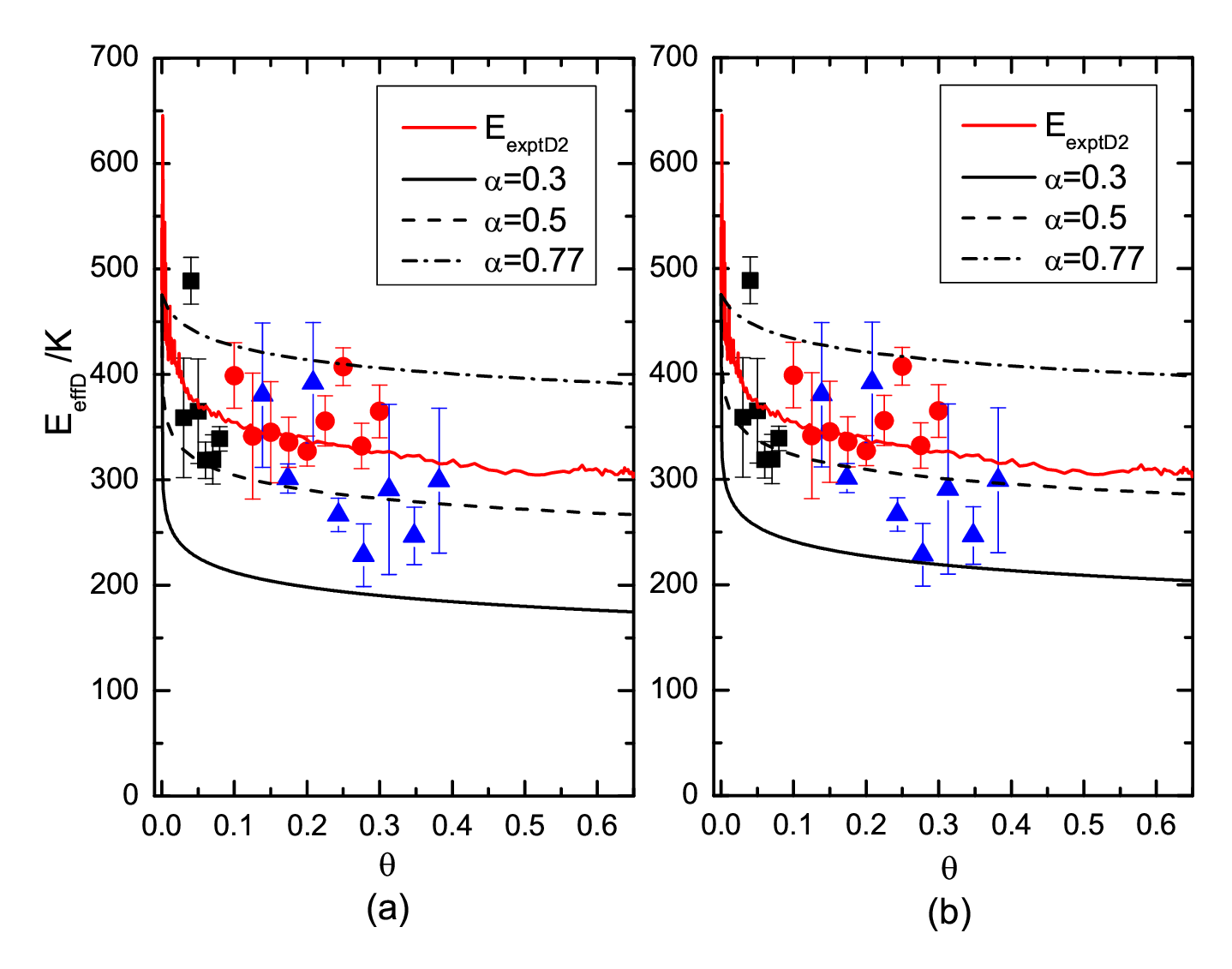}}
\caption{ $E_{effD}$ as a function of $\theta$ at fixed surface temperature $T$ = 20 K. Triangles, circles and squares represent desorption energies derived by the complete analysis, $E_{exptD}$~\citep{Tsuge2019}. Panel (a): $E_{D {H_{2}}}$ = 23 K, panel (b): $E_{D {H_{2}}}$= 73 K.}
\label{fig1}
\end{figure}

Fig.~\ref{fig1} shows $E_{effD}$ as a function of the relative coverage $\theta$. The surface temperature is fixed at $T$ = 20 K. We can see that $E_{effD}$ qualitatively agree with $E_{exptD2}$ and $E_{exptD}$ regardless of the values of 
$\alpha$ or $E_{D {H_{2}}}$ used in the calculations. The curves for $E_{effD}(\theta)$ are similar to these for $E_{exptD2}(\theta)$. The effective desorption energy is  more than 464 K (40 meV ) at very small relative coverage ($\theta\le 0.01$). At lower coverage ($\theta < 0.04$), $E_{effD}$ drops
quickly as  $\theta$ increases. But at higher coverage ($\theta > 0.04$), $E_{effD}$ only slightly decreases as $\theta$ becomes larger.  

Fig. \ref{fig1} also shows that $E_{effD}$ increases with increasing $\alpha$. This phenomenon can be explained as the follows. As $\alpha$ increases, H$_2$ molecules diffuse 
slower on surfaces. As a result, they are less likely to encounter each other. Thus, less encounter desorption events occur as $\alpha$ becomes larger.

Comparing panels (a) and (b) in Fig. \ref{fig1}, we can see that $E_{effD}$ also increases with increasing $E_{D {H_{2}}}$. This phenomenon is also easy to explain. Hydrogen molecules are less likely to desorb when they encounter with each other as $E_{D {H_{2}}}$ becomes larger, so less desorption events occur as $E_{D {H_{2}}}$ becomes larger.

\begin{figure}
\centering
\resizebox{8.5cm}{6.8cm}{\includegraphics{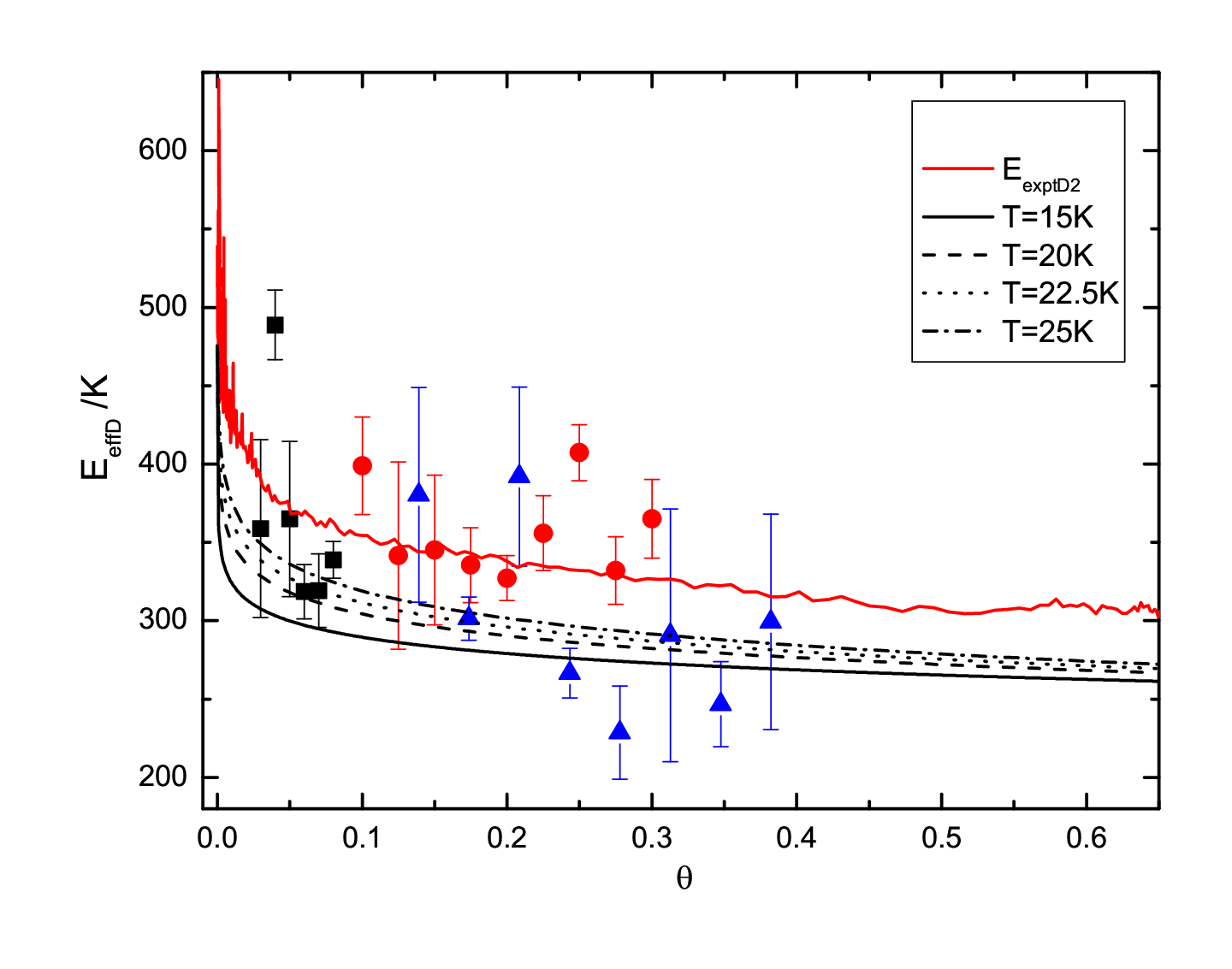}}
\caption{ $E_{effD}$ as a function of $\theta$ at various surface temperatures. $\alpha=0.5$, $E_{D {H_{2}}}$=23 K. Triangles, circles and squares represent desorption energies derived by the complete analysis, $E_{exptD}$~\citep{Tsuge2019}.
 }
\label{fig2}
\end{figure}

Fig. \ref{fig2} shows how surface temperatures affect $E_{effD}$. We fixed $\alpha$ and $E_{D {H_{2}}}$ at 0.5 and 23 K respectively in the calculations. Overall, $E_{effD}$ always increases as the surface temperature $T$ increases. On the other hand, $E_{effD}$ at all surface temperatures is well below $E_{exptD2}$.

Fig. \ref{fig2} clearly shows that $E_{effD}$ becomes larger as $T$ becomes higher. We can analyze equation~(\ref{equ2}) 
to get the same conclusion. The effective desorption energy is 
 approximately $ E_D-kTexp(( E_{D_{DLC}} - E_{b_{DLC}} )/kT) \theta P_{des}$ at very small $\theta$. 
Because $E_D-E_b >0$, $ kTexp(( E_{D_{DLC}} - E_{b_{DLC}} )/kT)$ decreases as $T$ gets larger, 
thus, $E_{effD}$ increases as $T$ becomes larger. The dependence of $E_{effD}$ on $T$ is even more obvious 
at much larger $\theta$. 
We found $E_{effD} \approx E_{b_{DLC}} -kTln(\theta P_{des})$ if 
$ exp((E_{D_{DLC}} - E_{b_{DLC}})/kT) \theta P_{des} \gg 1 $.
So $E_{effD}$ linearly increases with $T$ if $\theta$ is large enough. Therefore, our analysis shows that $E_{effD}$ increases with increasing $T$ at both very small and large relative coverage. 

The above figures show that the discrepancy between $E_{effD}$ and $E_{exptD2}$ is larger than the variations in $E_{exptD}$. On the other hand, the similarity between curves for $E_{effD}$ and $E_{exptD2}$ suggests that we can vary the values of $\alpha$ and / or $E_{D {H_{2}}}$ in the calculations for better agreement between $E_{effD}$ and $E_{exptD2}$. For instance, $E_{effD}$ calculated using $\alpha=0.5$ is smaller than $E_{exptD2}$. Because $E_{effD}$ increases as $\alpha$ becomes larger, we can slightly increase $E_{effD}$, therefore, reduce the discrepancy between $E_{effD}$ and $E_{exptD2}$ by using $\alpha$ values that are slightly larger than 0.5 in the calculations. On the other hand, the values of $\alpha$ and $E_{D {H_{2}}}$ in the literature are just theoretical predictions. Therefore, these values may not represent the true values of $\alpha$ and $E_{D {H_{2}}}$ in the experiments. So, we recalculate the values of $\alpha$ and $E_{D {H_{2}}}$ based on the experimental data to reduce
the discrepancy. The obtained values of these two parameters are reported in the following two subsections.

\subsection{The values of $\bar{E}_{D {H_{2}}_b}$ } 

We assumed $\alpha_b=0.3$, 0.5 and 0.77 to calculate
$E_{D {H_{2}}_b}$. However, we found that equation~(\ref{equ3}) does not have a solution if $\alpha_b=0.77$ was assumed. Therefore, we only report results for $\alpha_b=0.3$ and 0.5 in Table~\ref{table1}. 

\begin{table}
\caption{The values of $\bar{E}_{D {H_{2}}_b}$}
\label{table1}
\resizebox{8.5cm}{!}{
\begin{tabular}{cccccc}
   \hline
   & $T$ = 15 K    & $T$ = 17.5K     & $T$ = 20 K & $T$ = 22.5 K & $T$ = 25 K\\
  \hline
   $\alpha_b=0.3$   &  245 K        & 239 K       &  233 K & 227 K  & 220 K \\
   $\alpha_b=0.5$     &  153 K        &  144 K   &  134 K & 124 K  & 114 K \\
  \hline
\end{tabular}}
\end{table}

Table~\ref{table1} shows that $\bar{E}_{D {H_{2}}_b}$ decreases as the surface temperature becomes larger regardless of the adopted $\alpha_b$ values. Moreover, these $\bar{E}_{D {H_{2}}_b}$ values are all higher than the available desorption energies of H$_2$ molecules on H$_2$ in the literature~\citep{Cuppen2007,Das2021}. On the other hand, we can see that as $\alpha_b$ becomes larger, $\bar{E}_{D {H_{2}}_b}$ becomes smaller.

\begin{figure}
\centering
\resizebox{8.5cm}{6.8cm}{\includegraphics{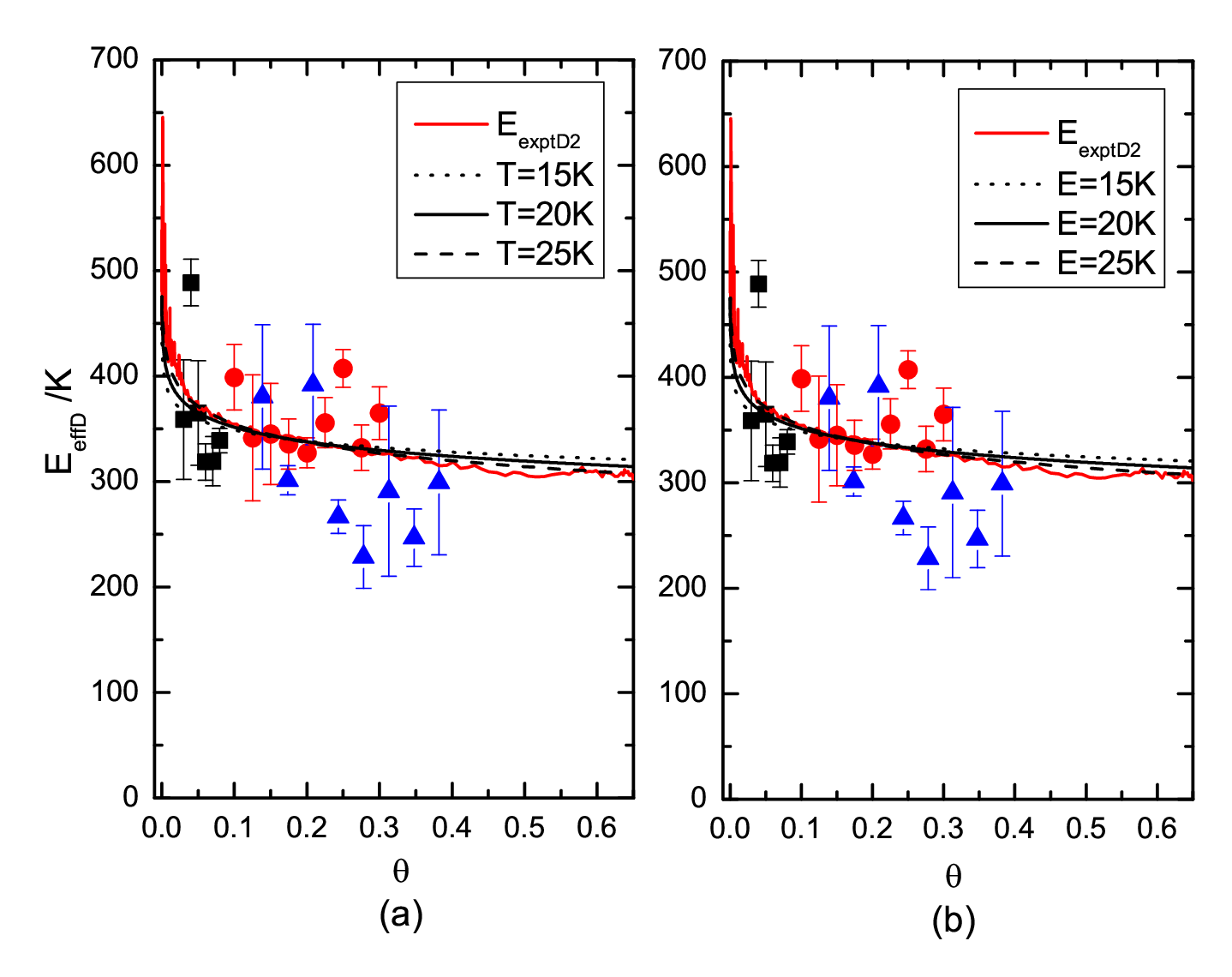}}
\caption{ $E_{effD}$ as a function of $\theta$. The updated values of $E_{D {H_{2}}}$ were used in the calculations, but $\alpha$ values are still these in the literature. The values of $E_{D {H_{2}}}$ and $\alpha$ used in the calculations depend on $T$. Triangles, circles and squares represent desorption energies derived by the complete analysis, $E_{exptD}$~\citep{Tsuge2019}. Panel (a): $\alpha=0.3$. $E_{D {H_{2}}}$ was set to be 245, 233 and 220 K for $T$ = 15, 20 and 25 K respectively. Panel (b): $\alpha=0.5$.  $E_{D {H_{2}}}$ was set to be 153, 134 and 114 K for $T$ = 15, 20 and 25 K respectively.}
\label{fig3}
\end{figure}

We calculated $E_{effD}(\theta)$ using the $\bar{E}_{D {H_{2}}_b}$ values listed in Table ~\ref{table1}. The values of $E_{D {H_{2}}}$ used in the calculations depends on the surface temperatures. Fig. \ref{fig3} shows the calculated $E_{effD}(\theta)$ as a function of $\theta$. Comparing panels (a) and (b) in Fig. \ref{fig3}, we found that $E_{effD}(\theta)$ agrees slightly better with experimental results if $\alpha=0.5$ is used in the calculations. But the difference is not significant, therefore, we do not distinguish these two panels in the following discussions. At lower coverage ($\theta<0.05$ ), $E_{effD}(\theta)$ is slightly smaller than $E_{exptD2}(\theta)$ for all the surface temperatures, but
at higher coverage, $E_{effD}(\theta)$ is larger than $E_{exptD2}$. Despite the small discrepancies, the differences between $E_{effD}(\theta)$ and $E_{exptD2}(\theta)$ are less than the variations in $E_{exptD}$. 
Moreover, $E_{effD}(\theta)$ agrees  better with $E_{exptD2}$ as $T$ increases from 15 K to 25 K. So, the newly calculated $E_{effD}(\theta)$ agrees well with experimental results regardless of $T$ and $\alpha$ used in the calculations. 

\begin{figure}
\centering
\resizebox{8.5cm}{6.8cm}{\includegraphics{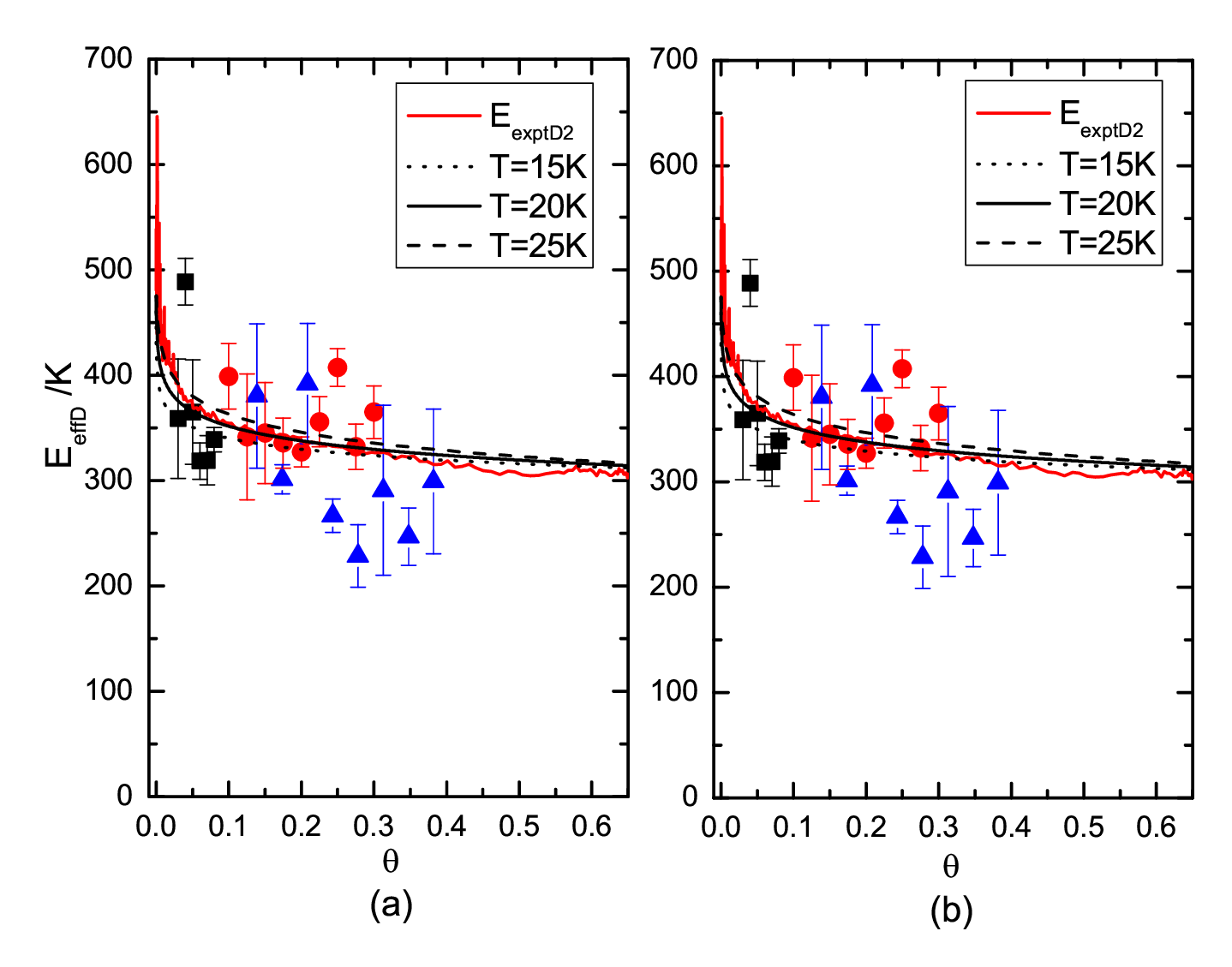}}
\caption{ $E_{effD}$ as a function of $\theta$. The updated values of $E_{D {H_{2}}}$ were used in the calculations, but $\alpha$ values are still in these in the literature. The values of $E_{D {H_{2}}}$ and $\alpha$ used in the calculations do not depend on $T$. Triangles, circles and squares represent desorption energies derived by the complete analysis, $E_{exptD}$~\citep{Tsuge2019}. Panel (a): $\alpha=0.3$, $E_{D {H_{2}}}$ = 233 K. Panel (b): $\alpha=0.5$,  $E_{D {H_{2}}}$ = 134 K.}
\label{fig4}
\end{figure}

We also calculated $E_{effD}(\theta)$ using $E_{D {H_{2}}}$ values that are independent of $T$. 
The obtained $E_{effD}(\theta)$ is shown as a function of $\theta$ in Fig. \ref{fig4}. 
Comparing panels (a) and (b) in Fig. \ref{fig4}, we found that $E_{effD}(\theta)$ calculated using $\alpha=0.3$ does not differ much from that calculated using $\alpha=0.5$.
At lower coverage ($\theta<0.05$ ), $E_{effD}(\theta)$ is slightly smaller than $E_{exptD2}(\theta)$ at $T$ = 15 K, however, the discrepancy between $E_{effD}$ and $E_{exptD2}$ almost disappears at $T$ = 25 K. At higher coverage ($\theta>0.3$ ), $E_{effD}(\theta)$ is always larger than $E_{exptD2}(\theta)$ regardless of $T$. On the other hand, the difference between  $E_{effD}(\theta)$ and $E_{exptD2}$ is also smaller than the variations in $E_{exptD}$. Therefore, $E_{effD}(\theta)$ calculated using the temperature independent $E_{D {H_{2}}}$ also fits in well with the experimental results.

\subsection{The values of $\bar{\alpha}_b$} 

Table~\ref{table2} shows our calculated $\bar{\alpha}_b$ at various surface temperatures. These $\bar{\alpha}_b$  values fall within the range of $\alpha$ in astrochemical models (0.3$\leq \alpha \leq 0.8$). The values of $\bar{\alpha}_b$ decreases as $T$ becomes larger regardless of the values of $E_{D {H_{2}}_b}$ used in the calculations. However, we can see that ($E_{D {H_{2}}_b}$(15 K) - $E_{D {H_{2}}_b}$(25 K))/$E_{D {H_{2}}_b}$(25 K) $<$ 0.1, which suggests that
differences of surface temperatures can only result in moderate variation of $\bar{\alpha}_b$. Finally, $\bar{\alpha}_b$ is larger if smaller $E_{D {H_{2}}_b}$ is used in the calculations.  

\begin{table}
\caption{The values of $\bar{\alpha}_b$}
\label{table2}
\resizebox{8.5cm}{!}{
\begin{tabular}{cccccc}
   \hline
             & $T$ = 15 K    & $T$ = 17.5K     & $T$ = 20 K & $T$ = 22.5 K & $T$ = 25 K\\
  \hline
  $E_{D {H_{2}}_b}$ =23 K  & 0.63 & 0.62 &  0.60 & 0.59  & 0.58 \\
  $E_{D {H_{2}}_b}$  =73 K & 0.59 & 0.58 &  0.57 & 0.55  & 0.54 \\
  \hline
\end{tabular}}
\end{table}

\begin{figure}
\centering
\resizebox{8.5cm}{6.8cm}{\includegraphics{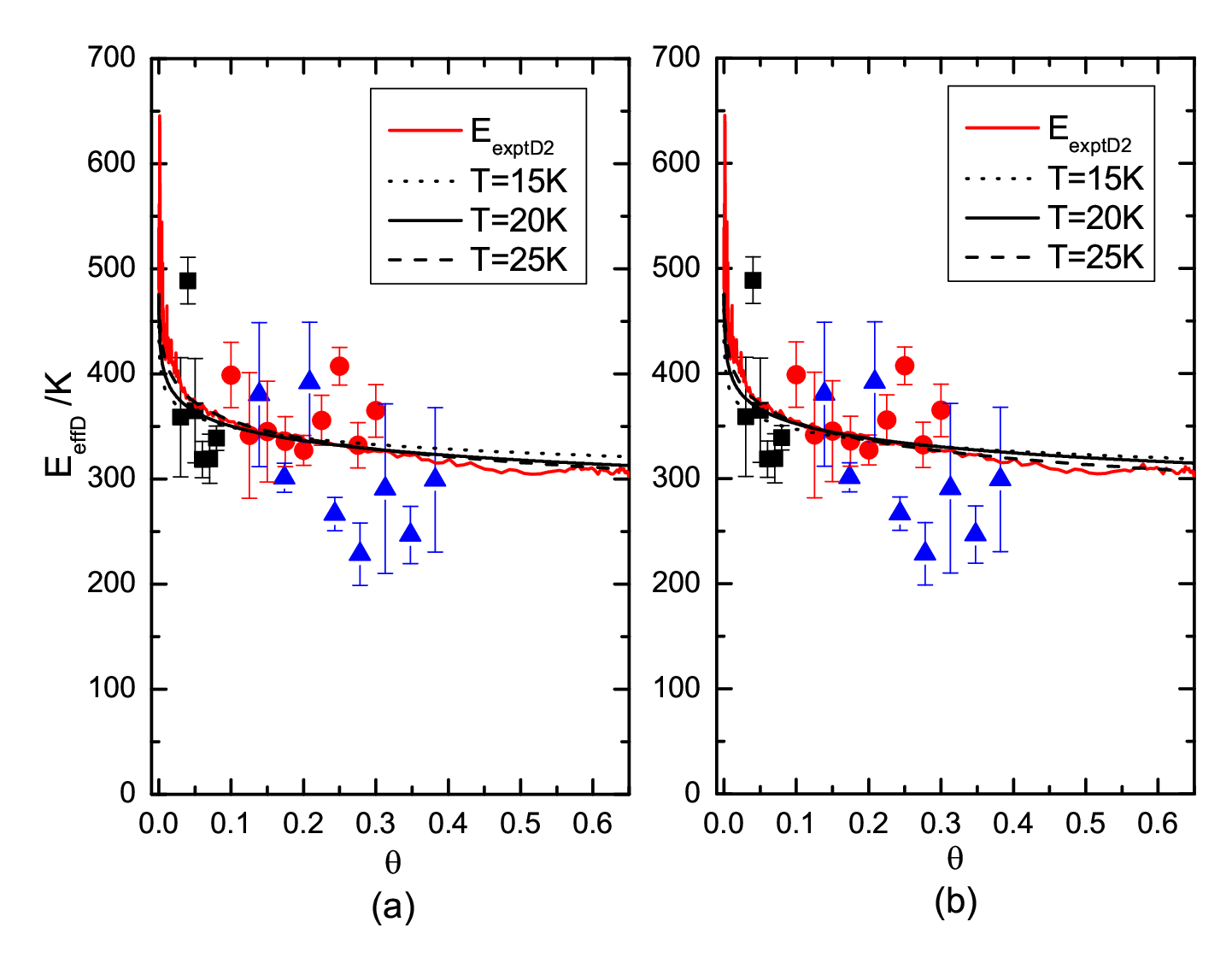}}
\caption{ $E_{effD}$ as a function of $\theta$. The updated values of $\alpha$ were used in the calculations, but $E_{D {H_{2}}}$ values used are still in these in the literature. The values of $\alpha$ used in the calculations depend on $T$. Triangles, circles and squares represent desorption energies derived by the complete analysis, $E_{exptD}$~\citep{Tsuge2019}. Panel (a): $E_{D {H_{2}}}$ = 23 K.  $\alpha$ was set to be 0.63, 0.6 and 0.58 for $T$ =15, 20 and 25 K respectively. Panel (b): $E_{D {H_{2}}}$ = 73 K.  $\alpha$ was set to be 0.59, 0.57 and 0.54 for $T$ =15, 20 and 25 K respectively.}
\label{fig5}
\end{figure}

We set $\alpha=\bar{\alpha}_b$ and $E_{D {H_{2}}}= E_{D {H_{2}}_b}$ to calculate $E_{effD}$. The values of $\bar{\alpha}_b$ depend on $T$. Fig. \ref{fig5} shows the calculated $E_{effD}$ as a function of $\theta$. Comparing the two panels in Fig. \ref{fig5}, we found that
$E_{effD}$ calculated using $E_{D {H_{2}}}~=~23$ K is almost the same as that using $E_{D {H_{2}}}~=~73$ K. At lower coverage ($\theta<0.05$), $E_{effD}$ is only slightly smaller than $E_{exptD2}$, but it becomes slightly larger than $E_{exptD2}$ at higher coverage. The discrepancy, however, is smaller than the variations in $E_{exptD}$. Therefore, $E_{effD}$ agrees well with $E_{exptD}$ and $E_{exptD2}$.

\begin{figure}
\centering
\resizebox{8.5cm}{6.8cm}{\includegraphics{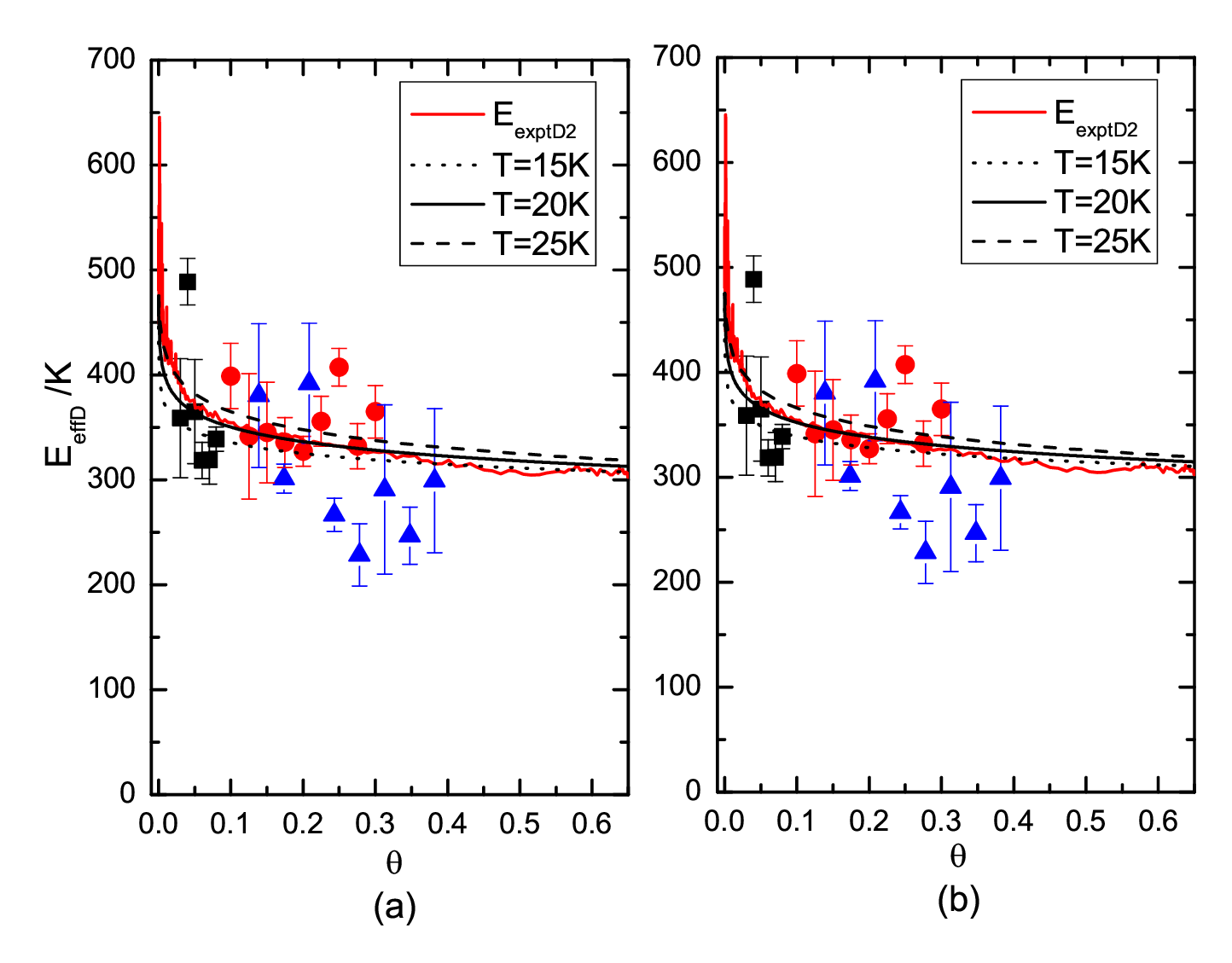}}
\caption{ $E_{effD}$ as a function of $\theta$. The updated values of $\alpha$ were used in the calculations, but $E_{D {H_{2}}}$ values used are still in these in the literature. The values of $\alpha$ used in the calculations do not depend on $T$. Triangles, circles and squares represent desorption energies derived by the complete analysis, $E_{exptD}$~\citep{Tsuge2019}. Panel (a): $\alpha=0.6$, $E_{D {H_{2}}}$ = 23 K. Panel (b): $\alpha=0.57$,  $E_{D {H_{2}}}$ = 73 K.}
\label{fig6}
\end{figure}

Good agreement between the calculated $E_{effD}$ and $E_{exptD}$ ($E_{exptD2}$) can also be achieved as shown in Fig. \ref{fig6} if temperature independent $\alpha$ values were used. Although $E_{effD}$ may be slightly larger or smaller than $E_{exptD2}$ depending on $\theta$, the discrepancy is smaller than the variation in $E_{exptD}$ regardless of $T$ and the values of $E_{D {H_{2}}}$ used in the calculations.

\section{Summaries and Discussions}

Using the encounter desorption mechanism, we explained the coverage dependent H$_2$ desorption energy, which was experimentally measured by \citet{Tsuge2019}. We suggested an effective desorption energy to compare with the coverage dependent H$_2$ desorption energy measured in the experiments. The value of the  effective desorption energy depends on $\alpha$ and $E_{D {H_{2}}}$. Using the values of these parameters in the literature, we can only qualitatively explain the coverage dependent H$_2$ desorption energy measured by the experiments. We argue that the discrepancy is because of current poor knowledge about $\alpha$ and $E_{D {H_{2}}}$. So, we calculated $\bar{\alpha}_b$ and $\bar{E}_{D {H_{2}}_b}$ based on experimental data. 

The values of $\bar{\alpha}_b$ and $\bar{E}_{D {H_{2}}_b}$ vary according to the surface temperatures. However, the calculated effective desorption energies at 15, 20 and 25 K agree well with experimental results even  $\alpha$ (or $E_{D {H_{2}}}$) is fixed to be the value of $\bar{\alpha}_b$ ($\bar{E}_{D {H_{2}}_b}$) at 20 K. Because the effective desorption energy increases monotonically with increasing $T$, all the effective desorption energies at the surface temperature range 15 - 25 K agree well with the experimental results. Because noticeable H$_2$ desorption occurs at temperatures between 15 K and 25 K only in the experiments~\citep{Tsuge2019}, we conclude that the coverage dependent H$_2$ desorption energy measured by the experiments can indeed be quantitatively reproduced by our approach.

In addition to reproducing the experimental results, our approach suggested a new method to measure H$_2$ diffusion barriers on surfaces, which are not well known so far. Rigorous quantum chemical calculations show that $E_{D {H_{2}}}=~73$ K~\citep{Das2021}  while the values of $\alpha$ in the literature are just crude estimations. Therefore, we argue that $E_{D {H_{2}}}~=~73$ K should be close to its true value in the experiments while the values of $\alpha$ in the literature deviate more from its true value in the experiment. Since our calculated effective desorption energies agree well with the experimental results, we argue that $\bar{\alpha}_b$ at $T$ = 20 K could be the value of $\alpha$ in the experiment. Therefore, the H$_2$ diffusion barrier on DLC surfaces prepared by~\citet{Tsuge2019} could be $\bar{\alpha}_b\times 41~meV\sim 23~meV$ (267 K). It should be noted that  that the term "DLC" represents a class of carbonaceous solid, 
where sp2-hybridized carbon atoms dominate. Because the $\alpha$ values should depend on surface, the H$_2$ diffusion 
barrier obtained above can be a representative value; it is therefore interesting to determine $\alpha$ values 
by changing the ratio of sp2 and sp3 contents. In any case, this work demonstrated that 
the combination of laboratory experiments employing TPD and numerical simulations is a useful tool to determine $\alpha$ values and, therefore, the H$_2$ diffusion barrier for various types of surface relevant to astronomical conditions.

Finally, we comment on the validity of ED mechanism since our approach is based on it. The repulsive interaction between neighboring adsorbate molecules were believed to be the reason for the decrease of desorption energy as the coverage of adsorbates increases~\citep{Wong2019}. For simplicity, we assume the repulsive interaction between adsorbate molecules is significant only when they are very close to each other, i.e., in the same binding sites. Due to the repulsive interaction, the desorption energy of adsorbates in the same binding sites should decrease, which is equivalent to reduction of desorption energy of surface species when they encounter in the ED mechanism. Therefore, the ED mechanism can be viewed as a simplification of the more complicated repulsive short-range interaction between adsorbates.  

\section*{Acknowledgements}
The research was funded by The National Natural Science Foundation of China under grants 12173023, 11973075, 11973099 and 12203091.
We thank Eric Herbst for helpful discussions.  We thank our referee for careful reading the manuscript and providing 
useful comments to improve the quality of the manuscript.
\section*{Data Availability}
No new data were generated or analysed in support of this research.











\bsp	
\label{lastpage}
\end{document}